\documentclass[fleqn,usenatbib]{mnras}

\usepackage{newtxtext,newtxmath}

\usepackage[T1]{fontenc}

\DeclareRobustCommand{\VAN}[3]{#2}
\let\VANthebibliography\thebibliography
\def\thebibliography{\DeclareRobustCommand{\VAN}[3]{##3}\VANthebibliography}

\usepackage{newtxmath}

\usepackage{graphicx}	
\usepackage{amsmath}	
\usepackage{amssymb}	
\usepackage{soul}

\title[Borderline Hyperbolic Comet C/2021 O3 (PANSTARRS)]{Borderline hyperbolic comet C/2021 O3 (PANSTARRS) was fading as it approached the Sun}

\author[M. Evangelista-Santana et al.]{M. Evangelista-Santana,$^{1}$\thanks{E-mail: marcalsantana@on.br}
M. De Pr\'a,$^{2}$
J. M. Carvano,$^{1}$
C.~de~la~Fuente~Marcos,$^{3}$
R. de la Fuente Marcos,$^{4}$
\newauthor
M R. Alarcon,$^{5}$
J. Licandro,$^{5,6}$
D. Lazzaro,$^{1}$
J. Michimani,$^{1}$
W. Pereira,$^{1}$
E. Rondón,$^{1}$
F. Monteiro,$^{1}$
\newauthor
P. Arcoverde,$^{1}$
T. Corr\^{e}a,$^{1}$
T. Rodrigues,$^{1}$
C. Paganini-Martins$^7$
\\
        $^1$Observat\'orio Nacional, rua Gal. Jos\'e Cristino 77, CEP 20921-400, Rio de Janeiro, Brazil\\
        $^2$University of Central Florida, Orlando, EUA\\
        $^3$Universidad Complutense de Madrid,
             Ciudad Universitaria, E-28040 Madrid, Spain \\
        $^4$AEGORA Research Group,
             Facultad de Ciencias Matem\'aticas,
             Universidad Complutense de Madrid,
             Ciudad Universitaria, E-28040 Madrid, Spain \\
$^{5}$Universidad de La Laguna, La Laguna, Spain\\
$^{6}$Instituto de Astrof\'{\i}sica de Canarias, Tenerife, Spain\\
$^{7}$Universidade Federal de Sergipe (UFS), Sergipe, Brazil \\
}
\date{Accepted XXX. Received YYY; in original form ZZZ}

\pubyear{2023}

\begin{document}
\label{firstpage}
\pagerange{\pageref{firstpage}--\pageref{lastpage}}
\maketitle

\begin{abstract}
We present an observational and numerical study of the borderline hyperbolic comet C/2021~O3 (PANSTARRS) performed during its recent passage through the inner Solar system. Our observations were carried out at OASI and SOAR between 2021 October and 2022 January, and reveal a low level of activity relative to which was measured for other long-period comets. In addition, we observed a decrease in brightness as the comet got closer to the Sun. Our photometric data, obtained as C/2021 O3 approached perihelion on 2022 April 21, show that the comet was much less active than what is usually expected in the cases of long-period comets, with $Af\rho$ values more in line with those of short-period comets (specifically, the Jupiter Family Comets). On the other hand, the observed increase in the value of the spectral slope as the amount of dust in the coma decreased could indicate that the smaller dust particles were being dispersed from the coma by radiation pressure faster than they were injected by possible sublimation jets. The analysis of its orbital evolution suggests that C/2021~O3 could be a dynamically old comet, or perhaps a new one masquerading as a dynamically old comet, with a likely origin in the Solar system. 
\end{abstract}


    

\section{Introduction}

   Until the beginning of the 21st century, the small Solar system bodies were roughly separated into two main groups: asteroids and comets. From the point of view of celestial mechanics, a classical dynamical criterion based on the Tisserand parameter (with Jupiter, $T_{\rm J}$, see e.g. \citealt{2000ssd..book.....M}), is often used to separate comets from asteroids, with comets presenting $T_{\rm J} < 3$ and asteroids, $T_{\rm J} > 3$. This parameter is a modified form of the Jacobi Integral, written in the inertial sidereal frame of reference; therefore, it is a constant of the circular restricted three-body problem. 
   
   Observationally, the cometary group is comprised of objects that present diffuse material surrounding a nucleus, while asteroids appear as a point of light. At the end of the previous century, the general consensus was that comets were made of ice-rich material and refractory solids, which would sublimate as they approach the Sun, while asteroids were composed of complexes of refractory minerals, with little to no volatile content (see e.g. \citealt{1993A&ARv...5...37F}). Despite this broad consensus, the existence of some overlap between the two groups and that members of one group may evolve into another has been suspected from some time (see e.g. \citealt{1971NASSP.267..413M,1980Natur.286...10H,1982come.coll..665D,1989NASSP.498...55W}).
    
    Recent advances in the study of comets, showed that ice sublimation is not the only mechanism for mass loss and that comets may be less volatile-rich than previously thought. Additional evidence suggests that after several perihelion passages, comets can lose or cover their volatile content and become inactive, presenting an asteroidal aspect \citep{1996EM&P...72..185J,2004come.book..317M}. These studies have moved away from a strict bimodal classification between asteroids and comets, towards an asteroid-comet continuum, with no clear division between these objects \citep{2022MNRAS.513.3734H,2022arXiv220301397J,2022ApJ...925L..15M}.
    
    Comets are a transient population of objects, with dynamical lives shorter than the age of the Solar system. There are two main reservoirs that replenish the population of comets: the Trans-Neptunian belt, the source of the short-period comets \citep{1991Icar...92..185I}, and the Oort Cloud, source of long-period comets \citep{2015SSRv..197..191D,2017ApJ...845...27N,2019AJ....157..181V}. Common end-states for both populations include: collisions with the Sun and the planets, becoming asteroids or interplanetary boulders, developing hyperbolic orbits, being ejected from the Solar system, sustaining fragmentation and disintegration by tidal breakup (when the comet approaches the Roche radius of a planet or the Sun), experiencing sublimation erosion, and undergoing rotational instability (when outgassing exerts a torque capable of changing the spin). Nonetheless, it is commonly accepted that comets can become inactive throughout their dynamical lives, before reaching their possible end-states \citep{2022AJ....164..158J}. 
    
    However, the mechanisms that lead to dormancy or extinction of cometary activity are not entirely understood. Theoretical work suggests that a dust mantle could be formed on the comet's surface as a result of the reaccumulation of the materials ejected over successive perihelion passages \citep{1984Icar...60..476F,1988Icar...76..493S}. This material would blanket the object's volatile content and prevent the activity from being triggered, inserting the comet into a dormant state. In this scenario, comets could be reactivated by inner Solar system perihelion passages, collisions, and other mechanisms that could reach or expose the hidden volatiles. Another possibility is that all volatile content could be lost, and the comet becomes extinct for the rest of its dynamical life \citep{1989aste.conf..880W}. 
    
    Studying comets with low-activity as possible candidates to being in the transitional phase may reveal key insights into the actual details of how these mechanisms operate. In this work, we investigate the low-activity, long-period comet C/2021~O3 (PANSTARRS). This object was discovered on 2021~July~26 at 4.3~au from the Sun, and moving towards perihelion at 0.287~au (2022~April~21), by the Panoramic Survey Telescope and Rapid Response System \citep{2004SPIE.5489...11K,2013PASP..125..357D}. The object presents an eccentricity $>1.0$, being also classified as a hyperbolic object. At discovery time, the comet was described as a ``soft" object, suggesting the existence of a coma with no tail, which was later confirmed by subsequent observations.\footnote{\url{http://www.cbat.eps.harvard.edu/iau/cbet/005000/CBET005009.txt} \label{foot}} Observations conducted after the perihelion passage by the Lowell Discovery Telescope (LDT) on 2022~April~29 indicated that the comet had possibly disintegrated \citep{2022ATel15358....1Z}.
    
    Here, we present a study of the dynamical and physical properties of the borderline hyperbolic comet C/2021~O3, based on observations made between the object's discovery time, and before its perihelion passage. In Section \ref{evolution}, we show dynamical simulations to investigate the objects' dynamical history. Photometric and spectroscopic observations are described in Section \ref{observations_reduct}, while the analysis of these data is presented in Section \ref{results}. Finally, in Section \ref{conclusions}, we present a discussion, and the conclusions and final remarks.

\section{Past, present and future dynamical evolution}\label{evolution}

   The marginally hyperbolic comet C/2021~O3 (PANSTARRS) was first identified by R. Weryk on July 26, 2021, in images obtained with the
   Pan-STARRS1 1.8-m Ritchey-Chretien reflector at Haleakala, Hawaii
   \citep{2021CBET.5009....1G,2021MPEC....P...05M} \footnote{\url{http://www.cbat.eps.harvard.edu/iau/cbet/005000/CBET005009.txt}}$^{,}$ \footnote{\url{https://mpcweb1.cfa.harvard.edu/mpec/K21/K21P05.html}}.
   The comet kept brightening as it approached its computed perihelion on April 21, 2022, at 0.29~au from the Sun. However, close
   to the Sun, it probably experienced a partial disintegration event that led to such a significant change from its pre-perihelion 
   orbital motion that, on April 29, 2022, it was recovered about two arcminutes southeast from the position predicted by the orbit 
   determination in use (see Table~\ref{elements}, orbit determination referred to epoch JD 2459493.5) at that moment 
   \citep{2022ATel15358....1Z}. The data in Table~\ref{elements} were obtained from Jet Propulsion Laboratory's Solar System 
   Dynamics Group Small-Body Database (JPL's SSDG SBDB, \citealt{2015IAUGA..2256293G})\footnote{\url{https://ssd.jpl.nasa.gov/sbdb.cgi}}.

   Although these early post-perihelion observations were interpreted as corresponding to the cloud of debris resulting from the 
   full disintegration of C/2021~O3 \citep{2022ATel15358....1Z}, the surviving object was recovered by multiple observatories
   during May at magnitudes not too different from those reported prior to reaching perihelion. This finding underlines the fact that 
   C/2021~O3 was subjected to a partial disintegration event, and not a full disruption episode like the one experienced by e.g. comet 
   C/2010~X1 (Elenin), a dynamically new Oort cloud comet that did not survive its first perihelion passage
   \citep{2015AJ....149..133L,2016EM&P..117..101K}\footnote{\url{https://www.nasa.gov/mission_pages/asteroids/news/elenin20111025.html}}. 
   It is known that comets with very small perihelion distances may experience complete or partial disruptions. Partial
   disruptions can lead to a sudden change in orbital motion that is better described as an abrupt transition rather than the result of 
   non-gravitational forces. 

   Average photometric magnitudes submitted by observers after May 19 were several magnitudes above the typical values reported 
   earlier and the values obtained in July were close to or above 20~mag. Consistently, the orbit determination was recalculated 
   using the observations of the surviving comet remnant (see Table~\ref{elements}, orbit determination referred to epoch JD 
   2459715.5). While the inbound heliocentric path was hyperbolic, its barycentric one was very eccentric but still elliptical 
   ($e<1$). This is often the case for dynamically old comets following very eccentric orbits that may have already completed 
   multiple perihelion passages. As for the path that C/2021~O3 is following after its partial disintegration event, both sets of 
   orbital elements (heliocentric and barycentric) are compatible with a hyperbolic trajectory, but perhaps the eccentricity is 
   not high enough to eject the comet into interstellar space like in the case of C/1980~E1 (Bowell) that reached the escape 
   velocity after experiencing a close encounter with Jupiter \citep{1982M&P....26..311B,2013RMxAA..49..111B}. 

%
%
   \begin{table*}
    \centering
    \fontsize{8}{12pt}\selectfont
    \tabcolsep 0.15truecm
    \caption{\label{elements}Values of the Heliocentric and Barycentric Keplerian orbital elements of C/2021~O3 (PANSTARRS) and
             their respective 1$\sigma$ uncertainties for the trajectories prior to and post disintegration.
             The orbit determination that reproduces the evolution prior to disintegration is based on 713 observations for a
             data-arc span of 190 days and it is referred to epoch JD 2459493.5 (2021-Oct-06.0) TDB (Barycentric Dynamical
             Time, J2000.0 ecliptic and equinox, solution date, 2022-May-23 12:46:32 PDT). The orbit determination applicable
             after disintegration is based on 64 observations for a data-arc span of 161 days and it is referred to as epoch JD
             2459715.5 (2022-May-16.0) TDB (Barycentric Dynamical Time, J2000.0 ecliptic and equinox, solution date,
             2022-Jul-19 01:12:05 PDT). Both orbit determinations were computed by D.~Farnocchia. The periapsis distance, for a
             heliocentric orbit, is the smallest distance between the object and the Sun; for a barycentric orbit, it is the 
             least distance between the object and the barycentre of the Solar system. This remark also applies to the time of 
             periapsis passage. Source: JPL's SBDB
            }
    \begin{tabular}{lccccc}
     \hline
                                                        &   & \multicolumn{2}{c}{prior to disintegration} & \multicolumn{2}{c}{after disintegration}    \\
      Orbital parameter                                 &   & Heliocentric            & Barycentric  & Heliocentric            & Barycentric \\
     \hline
      Eccentricity, $e$ & = &   1.000137$\pm$0.000002 &   0.999570   &   1.000077$\pm$0.000009 &   1.004408\\
      Periapsis distance, $q$ (au)                      & = &   0.287305$\pm$0.000002 &   0.284271   &   0.287362$\pm$0.000003 &   0.297149   \\
      Inclination, $i$ (\degr)                          & = &  56.7598$\pm$0.0004     &  56.6510     &  56.7937$\pm$0.0003     &  56.7254     \\
      Longitude of the ascending node, $\Omega$ (\degr) & = & 189.0349$\pm$0.0003     & 189.2005     & 189.0209$\pm$0.0005     & 188.5619     \\
      Argument of perihelion, $\omega$ (\degr)          & = & 299.9813$\pm$0.0003     & 300.0735     & 299.987 7$\pm$ 0.0003     & 301.0775     \\
      Time of periapsis passage, $T_{\rm a}$ (JD)       & = & 2459690.5409$\pm$0.0007 & 2459689.7972 & 2459690.5455$\pm$0.0002 & 2459690.3077 \\
     \hline
    \end{tabular}
   \end{table*}
%
%
   Both orbit determinations in Table~\ref{elements} were computed without having to resource to non-gravitational accelerations 
   that are caused by outgassing (see e.g. \citealt{1981AREPS...9..113S}). This does not mean that outgassing was not taking place 
   in this case. Non-gravitational accelerations are detected by analysing astrometric data (see e.g. \citealt{1973AJ.....78..211M}). 
   If the overall non-gravitational force on a small body resulting from directional mass loss is negligible or if the available 
   astrometry is not precise enough to assign a statistically meaningful contribution to the non-gravitational force, then a 
   standard orbit determination could be sufficient to reproduce past and present observations of the small body under study and 
   also to predict its future short-term orbital evolution. On the other hand, the orbital evolution of the nuclei of fast comets with rotation 
   periods of a few hours is nearly unaffected by outgassing (see e.g. \citealt{2004come.book..281S}).
   
   As the orbit determinations in Table~\ref{elements} are affected by some uncertainty and a relatively close encounter with the
   Sun played a major role in the orbital evolution of this object near perihelion, we adopted a statistical approach analysing the 
   results of the integration of a large sample of orbits and focusing on how C/2021~O3 arrived to the path that followed prior to 
   its partial disintegration and how it will evolve after that. Our $N$-body simulations were performed using Aarseth's 
   implementation of the Hermite integrator \citep{2003gnbs.book.....A}. His direct $N$-body code is publicly available from the 
   website of the Institute of Astronomy of the University of Cambridge.\footnote{\url{http://www.ast.cam.ac.uk/~sverre/web/pages/nbody.htm}}
   Most of the input data used in our calculations were retrieved from JPL's \textsc{horizons}\footnote{\url{https://ssd.jpl.nasa.gov/?horizons}}
   ephemeris system \citep{GY99}. The \textsc{horizons} ephemeris system was recently updated, replacing the DE430/431 planetary 
   ephemeris, used since 2013, with the new DE440/441 solution \citep{2021AJ....161..105P}. JPL's data retrieval was carried out 
   using the tools provided by the \textsc{Python} package \textsc{Astroquery} \citep{2019AJ....157...98G}. Additional details of 
   the simulations discussed here can be found in \citet{2012MNRAS.427..728D}.

   In order to understand the possible origin of C/2021~O3 and its future evolution better, we performed integrations backward 
   and forward in time applying the Monte Carlo using the Covariance Matrix (MCCM) methodology described by 
   \citet{2015MNRAS.453.1288D} to generate control or clone orbits based on the nominal orbit determination (see 
   Table~\ref{elements}) but adding random noise on each orbital element by making use of the covariance matrix. Covariance 
   matrices were retrieved from JPL's SSDG SBDB using the \textsc{Python} package \textsc{Astroquery} and its \textsc{SBDBClass} 
   class.\footnote{\url{https://astroquery.readthedocs.io/en/latest/jplsbdb/jplsbdb.html}} The MCCM technique was used to generate 
   initial positions and velocities for 10$^{3}$ control orbits that were evolved dynamically using the direct $N$-body code.

\begin{figure}
    \centering
    \includegraphics[width=\linewidth]{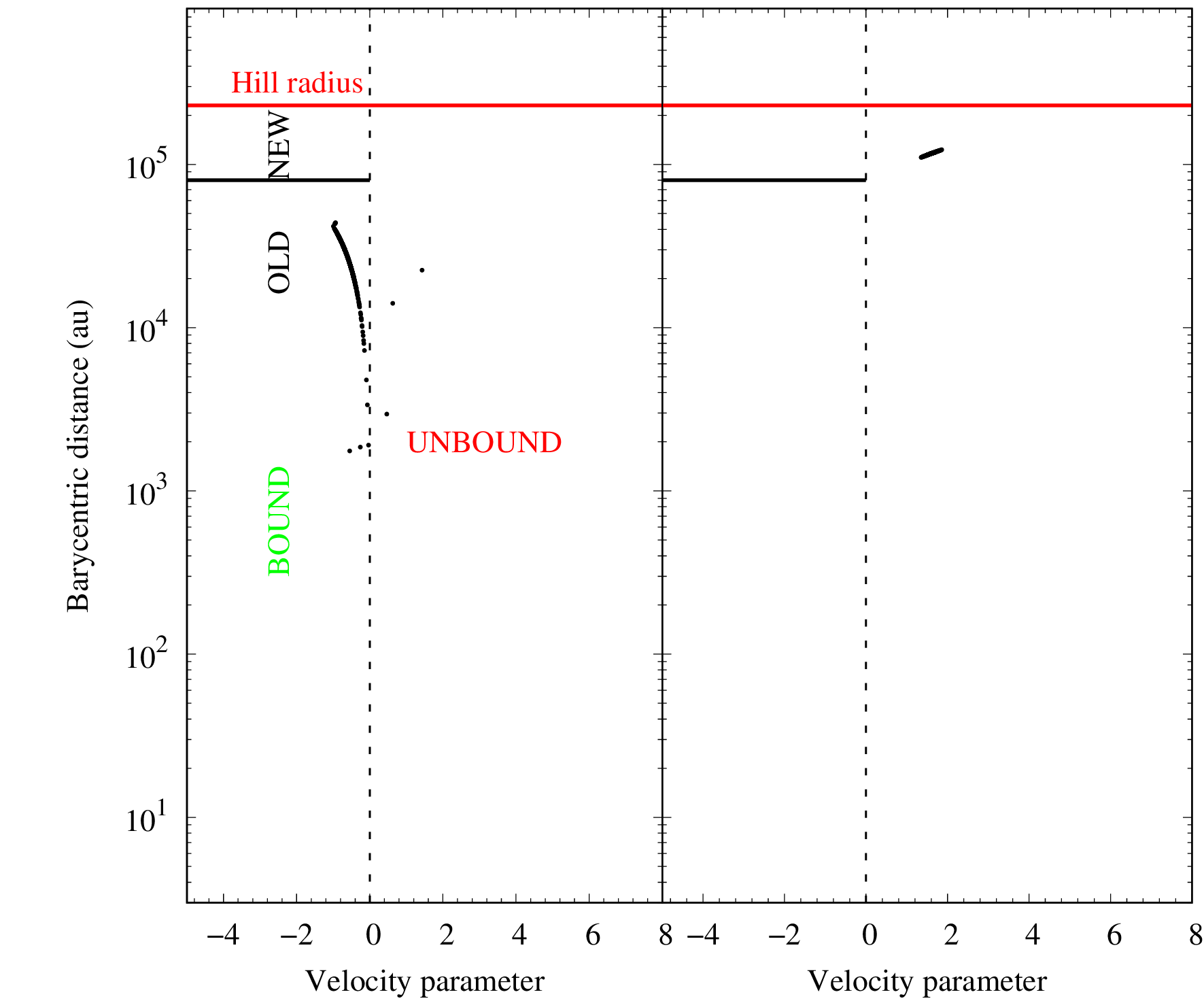}
    \caption{Values of the barycentric distance as a function of the velocity parameter 1.5~Myr into the past (left-hand side 
             panel) and 1.5~Myr into the future (right-hand side panel) for 10$^{3}$ control orbits of C/2021~O3 (PANSTARRS) generated using 
             the MCCM approach and based on the orbit determination referred to epoch JD 2459493.5 in Table~\ref{elements}. The 
             velocity parameter is the difference between the barycentric and escape velocities at the computed barycentric 
             distance in units of the escape velocity. Positive values of the velocity parameter are associated with control 
             orbits that could be the result of capture (left-hand side panel) or lead to ejection (right-hand side panel). The 
             thick black line corresponds to the aphelion distance ---$a \ (1 + e)$, limiting case $e=1$--- that defines the 
             domain of dynamically old comets with {$a^{-1}>2.5\times10^{-5}$~au$^{-1}$} (see \citealt{2017MNRAS.472.4634K}); the 
             thick red line signals the radius of the Hill sphere of the Solar system (see e.g. \citealt{1965SvA.....8..787C}).
            }
    \label{originC2021O3}
   \end{figure}

   Figure~\ref{originC2021O3} shows the result of the past and future evolution of C/2021~O3 according to the orbit determination 
   referred to epoch JD 2459493.5 in Table~\ref{elements}. Most control orbits led to barycentric distances with values below the 
   aphelion distance that defines the domain of dynamically old Oort cloud comets (see \citealt{2017MNRAS.472.4634K}) as shown in 
   the left-hand side panel of Fig.~\ref{originC2021O3}. The probability of having this comet captured from interstellar space 
   during the last 1.5~Myr is 0.005$\pm$0.009 (average and standard deviation of 10$^{3}$ experiments). The most straightforward 
   interpretation of these results is that C/2021~O3 could be a dynamically old comet (but see Sect.~5), with a likely origin in the Solar system.
   In absence of the partial disintegration event, the comet was certainly going to be ejected from the Solar system as shown in
   the right-hand side panel of Fig.~\ref{originC2021O3}. The flyby with the Sun was sufficiently close to provide enough 
   acceleration to escape the Solar system and reach interstellar space at low relative velocity with respect to the barycentre of 
   the Solar system.
    
%
%
%
%

\begin{figure}
    \centering
    \includegraphics[width=\linewidth]{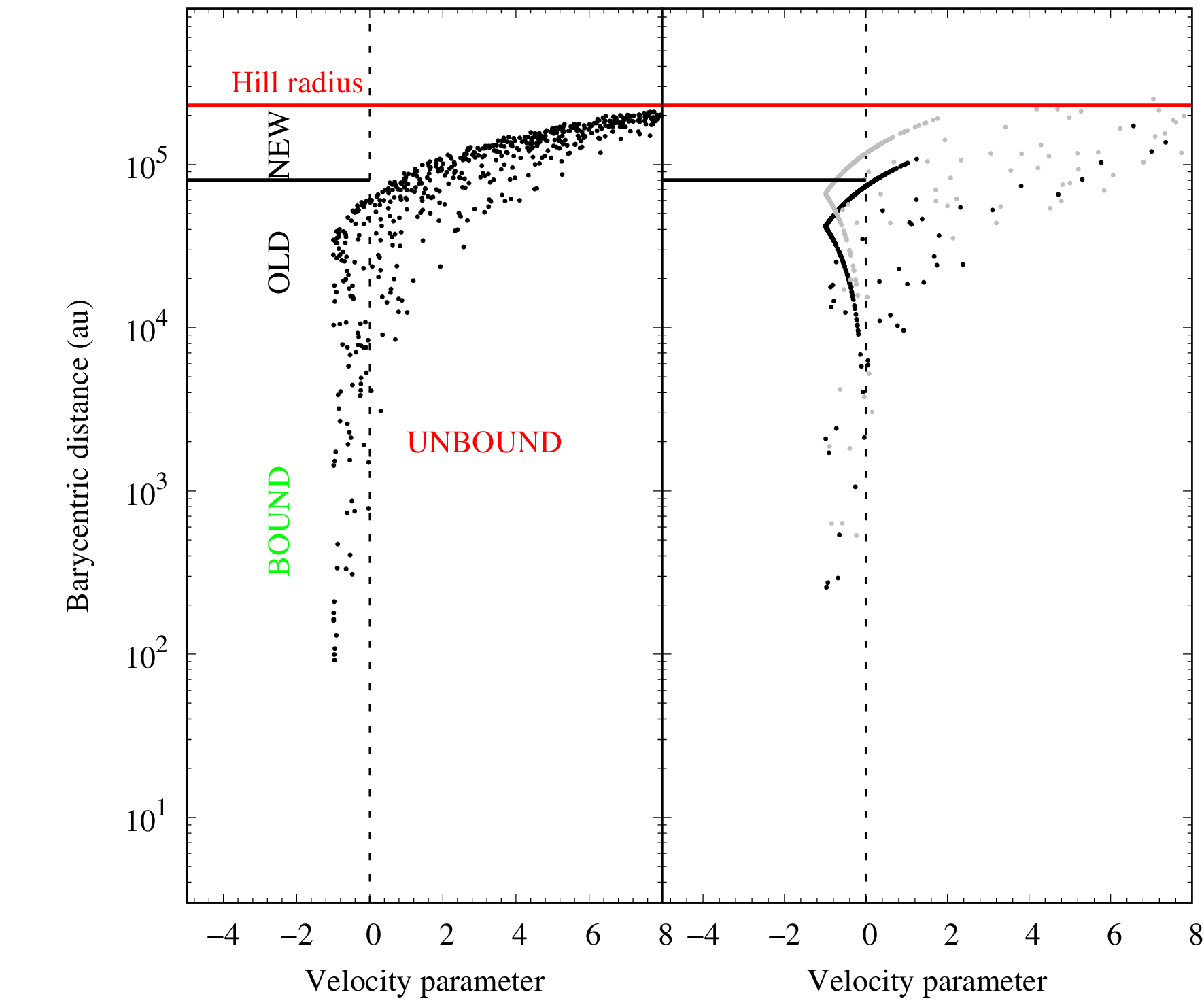}
    \caption{Values of the barycentric distance as a function of the velocity parameter 1.5~Myr into the past (left-hand side 
             panel) and 1.5~Myr (black) and 3~Myr (grey) into the future (right-hand side panel) for 10$^{3}$ control orbits of 
             C/2021~O3 (PANSTARRS) generated using the MCCM approach and based on the orbit determination referred to epoch JD 2459715.5 in 
             Table~\ref{elements}. The velocity parameter is the difference between the barycentric and escape velocities at the 
             computed barycentric distance in units of the escape velocity. Positive values of the velocity parameter are 
             associated with control orbits that could be the result of capture (left-hand side panel) or lead to ejection 
             (right-hand side panel). The thick black line corresponds to the aphelion distance ---$a \ (1 + e)$, limiting case 
             $e=1$--- that defines the domain of dynamically old comets with {$a^{-1}>2.5\times10^{-5}$~au$^{-1}$} (see 
             \citealt{2017MNRAS.472.4634K}); the thick red line signals the radius of the Hill sphere of the Solar system (see 
             e.g. \citealt{1965SvA.....8..787C}).
            }
    \label{futureC2021O3}
   \end{figure}

   Figure~\ref{futureC2021O3} shows the result of the past and future evolution of C/2021~O3 according to the orbit determination
   referred to epoch JD 2459715.5 in Table~\ref{elements}. Although the past orbital evolution is included for completeness, we 
   will focus on its future (right-hand side panel) to understand how such objects, resulting from a partial cometary 
   disintegration, may return to the inner Solar system and eventually be observed. Our calculations show that the partial 
   disintegration event led to a reduced probability of immediate ejection from the Solar system. Figure~\ref{originC2021O3},
   right-hand side panel, shows that the outbound journey of the original C/2021~O3 was leading into interstellar space with a 
   probability of 1. In sharp contrast, the change in the orbital motion of C/2021~O3 induced by the partial disintegration event
   translates into a reduced probability of 0.24$\pm$0.10 for an ejection within 1.5~Myr and 0.39$\pm$0.10 for 3~Myr. Our results 
   suggest that C/2021~O3 will continue being a member of the population of dynamically old Oort cloud comets for some time but 
   that it will eventually escape the Solar system if it is not fully disrupted in subsequent perihelion passages. On the other 
   hand, a fragmentation episode producing a group of genetically related comets as described by \citet{2021AJ....162...70Y} can 
   not be excluded. This behaviour has been observed in multiple long-period comets, for example C/2018~F4 (PANSTARRS) that
   broke-up in August 2020, a few months after reaching perihelion on December 4, 2019 
   \citep{2019A&A...625A.133L,2019RNAAS...3..143D,2021LPI....52.1107T}.

%
%
%
%

   Figure~\ref{futureC2021O3}, left-hand side panel, provides a cautionary tale for studies attempting to identify former 
   interstellar objects among known Solar system small bodies. The probability of having an interstellar origin for the comet 
   remnant resulting from the partial disintegration event of C/2021~O3 is 0.78 (integrating backward in time); however, we know 
   that the original C/2021~O3 was most likely a native of the Solar system. Fragments of former near-Sun comets, may be a 
   reasonable dynamical match for an inbound interstellar object approaching at low relative velocity with respect to the 
   barycentre of the Solar system.

\section{Observations and data reduction}\label{observations_reduct}

    \begin{table*}
	\centering
	\begin{tabular}{lcccccr} 
		\hline
		Site &  Date & Data type & Filter/spectral range ($\mu$m) & $\Delta$ (au) & $r$ (au) & $\alpha$ (\degr) \\
		\hline
		SOAR & $2021-10-14$ & Spectrum & ($0.5-0.9$)                & 2.587 & 3.348 & 12.5\\
        SOAR & $2021-10-14$ & Image   & $r$-band                    & 2.587 & 3.348 & 12.5\\
        OASI & $2021-11-02$ & Images   & $g$-, $r$-, $i$-band       & 2.563 & 3.104 & 17.0\\
        OASI & $2021-11-03$ & Images   & $g$-, $r$-band             & 2.563 & 3.090 & 17.2\\
        OASI & $2021-11-05$ & Images   & $g$-, $r$-, $i$-, $z$-band & 2.565 & 3.064 & 17.4\\ 
        OASI & $2021-12-06$ & Images   & $g$-, $r$-, $i$-band       & 2.627 & 2.640 & 21.0\\
        OASI & $2021-12-07$ & Image   & $r$-band                    & 2.629 & 2.626 & 21.6\\
        OASI & $2021-12-09$ & Images   & $g$-, $r$-band             & 2.633 & 2.598 & 21.7\\
        OASI & $2022-01-04$ & Images   & $g$-, $r$-band             & 2.645 & 2.261 & 21.3\\
		\hline
	\end{tabular}
	\caption{All the observations of C/2021~O3 (PANSTARRS) obtained with the SOAR and OASI telescopes. For each entry, $\Delta$ is the geometric distance to Earth, $r$ is the heliocentric distance, and $\alpha$ is the phase angle at the time of the observations.}
    \label{observations}
    \end{table*}

    \subsection{Observations}
    The observations of C/2021~O3 (PANSTARRS) were obtained at the \textit{Observat\'{o}rio Astron\^{o}mico do Sert\~{a}o de Itaparica} (code Y28, OASI, Nova Itacuruba) in Brazil, with a 1.0~m f/8 telescope (Astro Optik, Germany) and a $2048\times2048$ pixel CCD FLI camera. This instrument produces images with a pixel scale of 0.3457~arcsec~pixel$^{-1}$. In this configuration, we used  $g$-, $r$-, $i$-, and $z$-band filters in the Sloan Digital Sky Survey (SDSS) system. More details on the avaliable instrumentation at OASI are given in \cite{2020PASP..132f5001R}.
    
    At the Southern Astrophysical Research (SOAR) 4-m telescope, we used the Goodman High Throughput Spectrograph (GHTS) to acquire the object's spectrum and images, both with Red Camera. For the spectroscopic observations we used a grating of 400~l/mm and a 3.2" slit positioned along the parallactic angle and no binning, producing a visible spectra covering the wavelength range of $0.5-0.9$~$\mu$m. For the imaging mode we used a 1$\times$1 binning, resulting in an effective circular field of view of 7.2 arcmin (diameter) and the image scale of $0.15$ arcsec pixel$^{-1}$. All observations were made using the $r$-band filter in the Sloan Digital Sky Survey (SDSS) system.

    \subsection{Data reduction}
    The data were reduced using the Image Reduction and Analysis Facility (IRAF)\footnote{IRAF is distributed by the National Optical Astronomy Observatories, which are operated by the Association of Universities for Research in Astronomy, Inc., under cooperative agreement with the National Science Foundation.} and custom \textsc{Python} routines. The data obtained from OASI were first processed through a custom pipeline that performs flat field and dark frame corrections, calculates the astrometric solution, identifies stars from the \textit{Gaia}~DR2 catalogue \citep{2018A&A...616A..17A}, and flags known asteroids. Fringing images for the $i$- and $z$-band filters were produced by combining the dithered images in each filter, after processing with a noise-reducing wavelet filter. For the SOAR photometric  data, flat field and bias corrections were performed using IRAF routines. The images acquired at each observing run were centred at the optocentre of the comet and then average-combined in order to improve the signal-to-noise ratio. Aperture photometry on the combined images was performed using the task DIGIPHOT from the APPHOT package included in IRAF. We used isolated stars with magnitudes in the SDSS system or photometric calibrations \citep{1996AJ....111.1748F} to this end.   
    
    The spectroscopic data obtained at SOAR were corrected for bias and divided by a normalized flat. We extracted the 2D spectrum of C/2021~O3 (PANSTARRS) and those of the solar analogues using IRAF's \textit{APALL} task, centering the comet optocentre at the middle of the aperture. The background sky was subtracted in a region close to the comet, free of the coma, tail, and background stars. In sequence, the spectrum of the comet and the solar analogues were 
    calibrated in wavelength using HgArNe lamps, and then the average of the combined individual spectra of each target was computed to increase the signal-to-noise ratio. We obtained the reflectance spectrum by dividing the one of the comet by each of the solar analogues spectra observed that night. As a final step, a resulting reflectance spectrum of C/2021~O3 was produced by averaging all the individual reflectance spectra. The journal of the observations is given in Table \ref{observations}.


\section{ Data Analysis}\label{results}

   We investigate the evolution of three properties of C/2021~O3 (PANSTARRS) as it approached the perihelion: the total magnitudes, dust-production rates, and spectral slopes. 

    \subsection{Magnitudes and dust production}
    \label{dust}
    
    The apparent magnitude of the comet was measured using fixed aperture at reference $\rho\simeq10,000$~km (centred at the optocentre of the comet) for the entire observation period. In sequence, we corrected the apparent magnitudes by heliocentric and geocentric distances (1, 1, $\alpha$) to infer the reduced magnitudes. The values obtained are shown in Table \ref{Tab_afho} and correspond to the data obtained pre-perihelion. The values of the calculated magnitude show an increase from 2021~October~14 to 2022~January~04 of 0.44$\pm$0.06 with decreasing heliocentric distance varying from 3.348~au to 2.261~au.
    
    To evaluate the dust activity during the observations, we measured the $Af\rho$ parameter as a proxy for dust production \citep{1984AJ.....89..579A}. This parameter is given in length units (cm). In general, an increase of the activity of the comet is expected as the heliocentric distance decreases although long-period comets tend to have higher $Af\rho$ values than the short-period comets \citep{2021P&SS..20605308G}. We calculated the values of $Af\rho$ as a function of the distance to the optocentre for all observed epochs. The values corresponding to a reference aperture of $\rho=10^4$ km are shown in Table~\ref{Tab_afho}, where we also include the date of the observations, the filter of reference, the magnitude in the filter, the values of the maximum of $Af\rho$, and the aperture in kilometers where this maximum was obtained.

\begin{figure}
    \centering
    \includegraphics[width=\columnwidth]{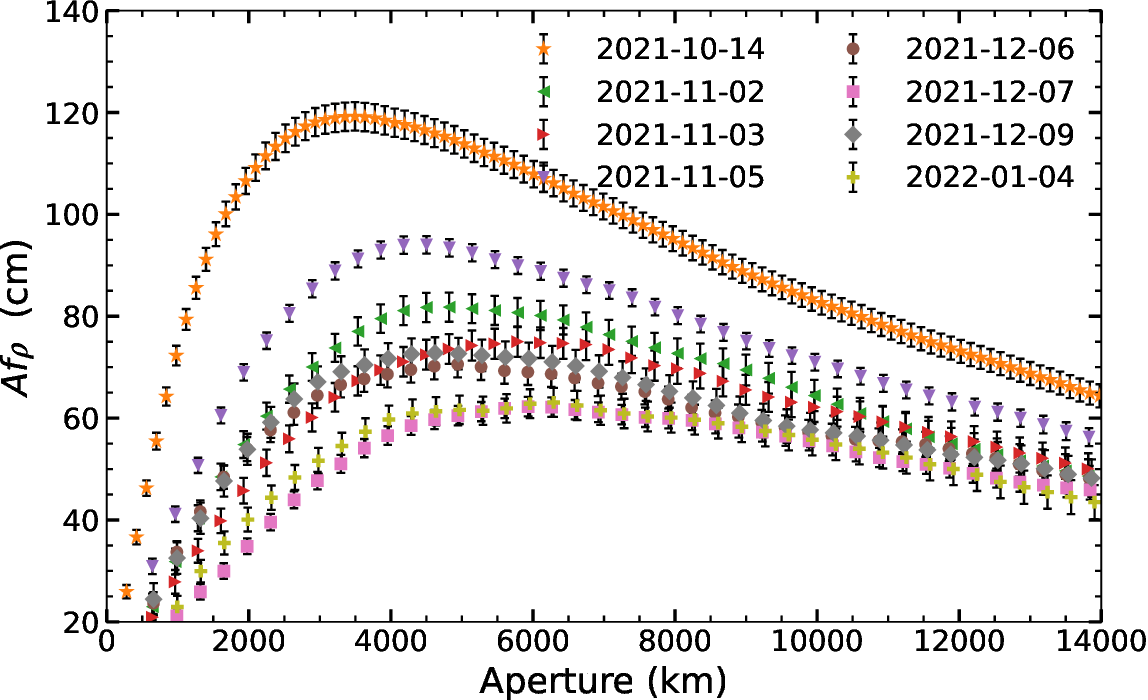}
    \caption{Variation of the dust production parameter $Af\rho$ with the aperture, centred at the optocentre of C/2021~O3 (PANSTARRS).\label{afrho_t}}
    \end{figure}

\begin{figure}
    \centering
    \includegraphics[width=\columnwidth]{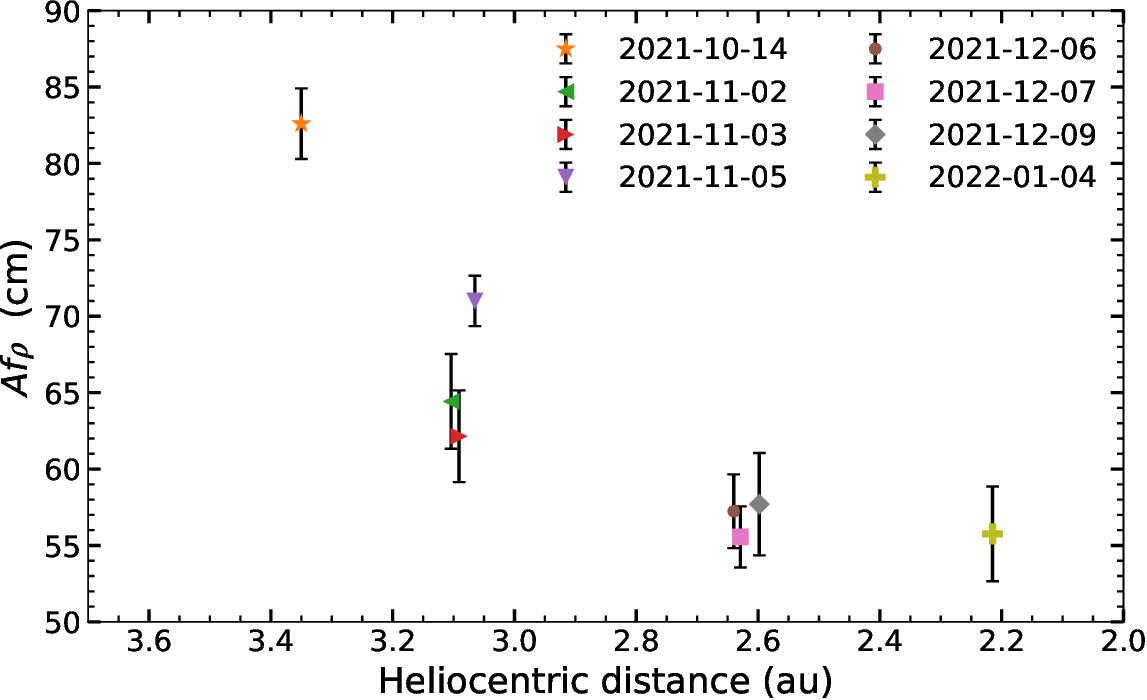}
    \caption{Evolution of the dust production rate of C/2021~O3 (PANSTARRS) at $\rho\simeq10,000$~km as the heliocentric distance changes.}
    \label{afrho}
    \end{figure}

    Figure \ref{afrho_t} shows the $Af\rho$ values obtained for different aperture radii for C/2021~O3 (PANSTARRS) during the observation period. In this figure, each curve corresponds to a date of observation and it is possible to compare the evolution of the activity and infer the dust distribution of the coma. From  Fig.~\ref{afrho_t}, we notice that the maximum $Af\rho$ values tend to decrease with decreasing heliocentric distance, with the exception of the values measured on November 5 and on December 9, 2021. Figure \ref{afrho} shows the $Af\rho$ values for the reference aperture as a function of the heliocentric distance. The data suggests a decrease in the dust production as the comet approached perihelion, with  a sudden increase on November 5, and stabilizing from December 6 onward.

    \begin{table*}
	\centering
	\caption{Af$\rho$ values for comet C/2021 O3 (PANSTARRS), derived for the SDSS filter system with the reference optical aperture of $\rho = 10^4$ km and centred at the optocentre.}
	\label{Tab_afho}
	\begin{tabular}{lcccccccr} 
		\hline
		Date & Filter & Magnitude (1,1,$\alpha$)& Af$\rho$ (cm) & Af$\rho_{\rm max}$ & $\rho_{\rm max}$ (km) \\
        & &($\rho = 10^4$ km) & ($\rho = 10^4$ km) & &  \\
		\hline
        2021-10-14 & $r$-band & 13.04 $\pm$0.03 &  82.64    $\pm$2.31 & 119.23 $\pm$2.80 & 3494.60 \\
        2021-11-02 & $r$-band & 13.32 $\pm$0.05 &  64.43	$\pm$3.10 & 81.80  $\pm$2.90 & 4819.60 \\
        2021-11-03 & $r$-band & 13.36 $\pm$0.02 &  62.15	$\pm$3.00 & 75.00  $\pm$3.00 & 5783.50 \\
        2021-11-05 & $r$-band & 13.21 $\pm$0.04 &  71.00	$\pm$1.65 & 94.00  $\pm$1.72 & 4501.80 \\
        2021-12-06 & $r$-band & 13.46 $\pm$0.04 &  57.24	$\pm$2.41 & 70.60  $\pm$1.60 & 4939.93 \\
        2021-12-07 & $r$-band & 13.49 $\pm$0.03 &  57.64	$\pm$3.30 & 64.21  $\pm$3.13 & 6921.17 \\
        2021-12-09 & $r$-band & 13.45 $\pm$0.06 &  57.11	$\pm$3.40 & 72.80  $\pm$3.10 & 4621.13 \\
        2022-01-04 & $r$-band & 13.48 $\pm$0.05 &  55.80    $\pm$3.10 & 63.10  $\pm$3.00 & 6290.60 \\
		\hline
	\end{tabular}
\end{table*}
                              
\begin{figure}
    \centering
    \includegraphics[width=\columnwidth]{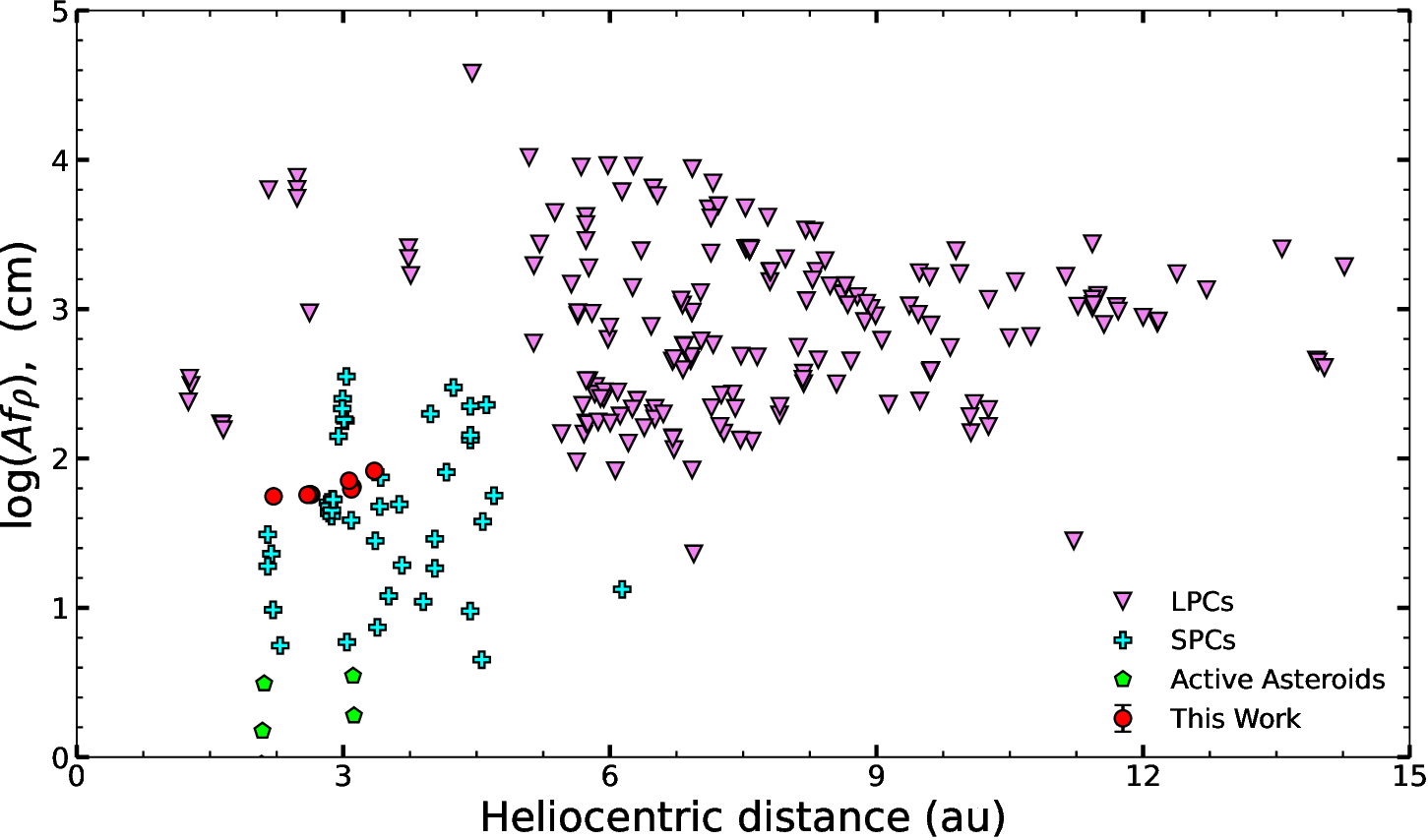}
    \caption{Values of $\log(Af\rho)$ computed for long-period comets (triangles), short-period comets (crosses), active asteroids (hexagons), and our data for C/2021~O3 (PANSTARRS) as circles. 
    \label{afrho_log}}
    \end{figure}

    In Fig.~\ref{afrho_log}, we show the  spatial distribution parameters, $\log(Af\rho)$, as a function of the heliocentric distance for a sample of long-period comets (LPCs, triangles), short-period comets (SPCs, crosses), active asteroids (hexagons) and our data (circles), along with the $Af\rho$ values derived for comet C/2021~O3 in this work. The values of $\log(Af\rho)$ are significantly larger for LPCs than for SPCs and active asteroids, but the values for comet C/2021~O3 are numerically closer to those of SPCs \citep{2009Icar..201..719M,2016AJ....152..220S,2016P&SS..132...23M,2020P&SS..18004779G,2019Icar..317...44B,2020AN....341..849B,2007MNRAS.381..713M,2020AstBu..75...31I}. 
   
\subsection{Spectral properties}

\begin{figure}
    \centering
    \includegraphics[width=\columnwidth]{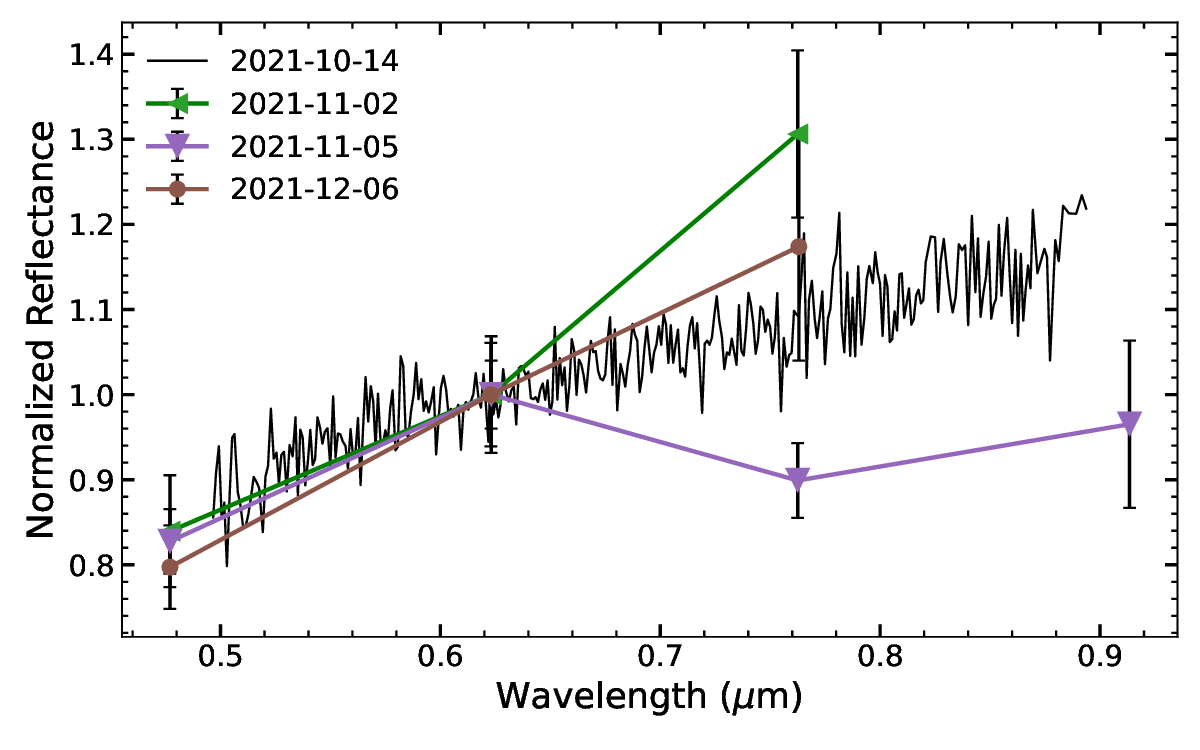}
    \caption{Reflectance data of hyperbolic comet C/2021~O3 (PANSTARRS) from low-resolution spectroscopy and SDSS colours.}
    \label{spectrum}
    \end{figure}

    Figure~\ref{spectrum} shows the low-resolution reflectance spectrum of C/2021~O3 (PANSTARRS) obtained at SOAR on October 14, 2021, along with the reflectance data derived from the OASI observing runs with three or more SDSS filters, on November 2 and 5. The magnitudes were calculated using the reference aperture and we considered solar colours from \citet{2018ApJS..236...47W} in order to calculate the reflectance from the SDSS colours. The SOAR spectrum has no signatures of the presence of aqueous-altered minerals. 

    Using the CANA package \citep{2018DPS....5031502D}, we obtained the taxonomic classification of C/2021~O3 from the low-resolution spectrum. Its spectrum  corresponds to that of a D-type asteroid; asteroids of this taxonomic type are common in the outer main belt and among the Jupiter Trojans \citep{2018Icar..311...35D}. On the other hand, cometary spectra are also affected by the size distribution of the dust in the coma \citep{2017MNRAS.468.1556R}.
    We also used the CANA package to calculate the spectral slope of the low-resolution spectrum of the comet. We compute the $S'$ in the wavelength range of $0.55-0.85$~$\mu$m following the definition in \cite{1996AJ....111..499L}. The fitting was normalized to the unit at 0.55~$\mu$m and $S'$ computed in units of $\%10^{-3}$\AA. The value obtained is $S'=7.32\pm$0.15~(\%$10^{-3}$\AA).

    The reflectance spectra of C/2021~O3 show a marked change in morphology as the dust production varies over time, but no clear trend can be seen from the four nights presented in Fig.~\ref{spectrum}: the spectrum initially becomes redder on November~3, but, on November~5, the comet's reflectance spectrum acquires an unusual shape, returning on December~6 to a D-type spectrum that is somewhat redder than the low-resolution spectrum of October~14. 

\begin{figure}
    \centering
    \includegraphics[width=\columnwidth]{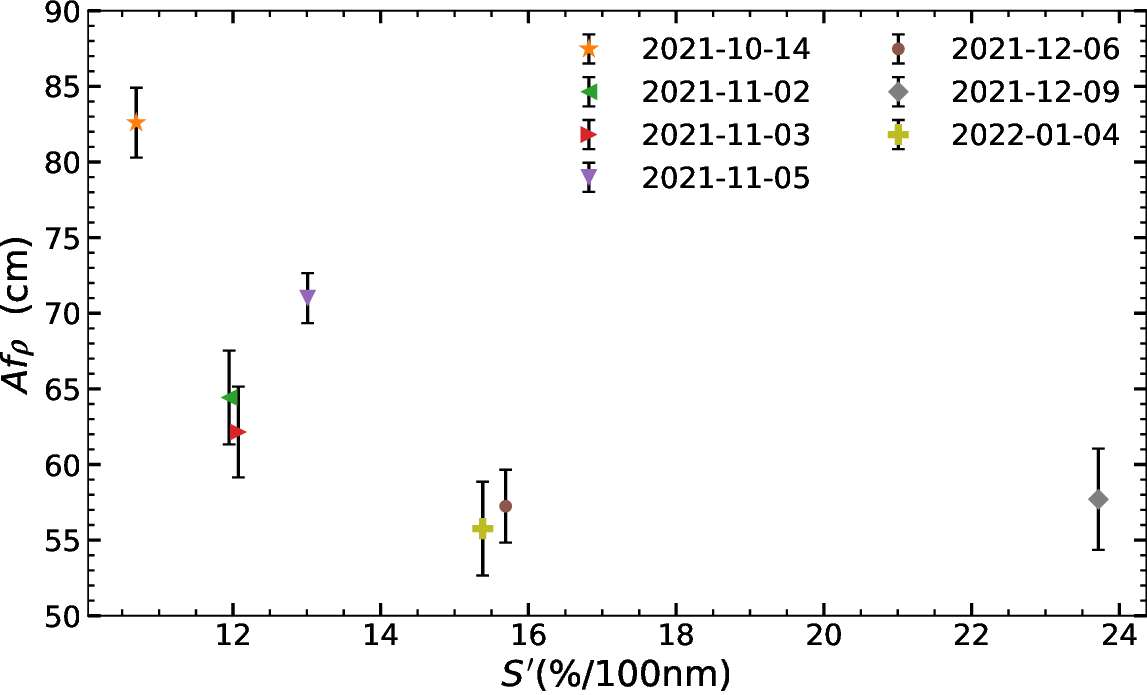}
    \caption{Evolution of the dust production parameter $Af\rho$ with the spectral slope for C/2021~O3 (PANSTARRS) as it approached perihelion.}
    \label{afrho_spectral}
    \end{figure}

    In order to better understand the relation between spectral slope and the dust content of the coma, we initially calculated the spectral gradient $S'$ considering only the reflectance in the $g$ and $r$ bands, which were observed on all nights. For the low-resolution spectrum, the spectral slope was calculated considering the wavelength range $0.477-0.623$~$\mu$m. Figure~\ref{afrho_spectral} shows a scatter plot of the values of spectral slope and $Af\rho$. Considering the values of $S'$ calculated from the $g$ and $r$ filters, the spectra initially becomes redder as the dust content in the coma decreases, but the correlation ceases to exist as $Af\rho$ stabilizes around $Af\rho\sim55$~cm after December 2021. A rather extreme value of $S'\approx24\%/100$~nm was measured on December~9, which is significantly higher than the values measured on December~6, 2021, and January 4, 2022, with similar values of $Af\rho$. Also, the initial trend of redder spectra with decreasing dust content is interrupted on November~5, which is the same night that is an outlier in the trend of decreasing $Af\rho$ with decreasing heliocentric distance shown in Fig.~\ref{afrho}, and also presented the unusual reflectance spectrum seen in Fig.~\ref{spectrum}.

    \section{Discussion and Conclusions}\label{conclusions}

    The marginally hyperbolic comet C/2021~O3 (PANSTARRS), as detailed in Sect.~\ref{evolution}, was initially identified in images obtained by PanSTARRS1 on July~26, 2021, at 4.3~au from the Sun and moving to reach perihelion at 0.29~au on April~21, 2022. However, post-perihelion observations made on April~29, 2022 and reported in \citet{2022ATel15358....1Z} showed that the observing team failed to locate any object at the predicted position and expected magnitude of C/2021~O3, which would place an upper limit for the dust content of $Af\rho<10$~cm. However, \citet{2022ATel15358....1Z} also detected a diffuse glow with a diameter of $\sim40"$ about 2' away from the predicted position, which seemed to be co-moving consistently with the original orbital solution for the comet.  \citet{2022ATel15358....1Z} interpreted this finding as an evidence of the disintegration of the nucleus. After this report was issued, multiple observatories detected a signal from the comet, suggesting a cometary split or a partial disintegration event, but not a full disruption. Subsequent observations showed a significant alteration in the orbital path of the surviving object with respect to the computed pre-perihelion orbit. 
     
    Our photometric data obtained as the comet approached perihelion showed that the comet was much less active than what is usually expected for long-period comets, with $Af\rho$ values more in line with those of short-period comets. Also, even if the observed integrated magnitude between the distance intervals (3.348, 2.261)~au from the Sun increased by 0.44$\pm0.06$~mag, the comet was in fact becoming less active as it approached the Sun, with decreasing values of $Af\rho$. The observed increase in the spectral slope as the amount of dust in the coma decreased could indicate that the smaller dust particles were being dispersed from the coma by radiation pressure faster than they were injected by sublimation jets. This interpretation is consistent with the fact that the sudden increase in $Af\rho$ observed on the outburst of November 5 resulted also in a significant change in the comet spectrum, with a marked increase in reflectance at lower wavelengths.  
     
    Considering C/2021~O3 as a comet that is dynamically old, as implied by the dynamical analysis (but see below), our observations suggest that the multiple previous passages through the inner Solar system had gradually depleted its volatile content or that the active regions were covered by an insulating layer, although this last option is less likely if we consider the comet to be small in size. The comet was therefore becoming inactive as it approached perihelion, and the fact that the comet was briefly lost post-perihelion \citep{2022ATel15358....1Z} would be consistent with a near complete dissipation of the coma. Both the recovery and subsequent change in orbit suggest that the comet was violently reactivated as the heating during perihelion penetrated the inner layers still rich in volatiles.
        
    Following the classification criteria discussed by \citet{2017MNRAS.472.4634K}, dynamical integrations based on the pre-perihelion orbital elements suggest that C/2021~O3 could be a dynamically old comet, with a likely origin in the Solar system. Stating that there is no need to invoke an interstellar source for this marginally hyperbolic small comet is strongly supported by the data. However, its degree of evolutionary maturation is far from clear. Although at face value, our results are consistent with C/2021~O3 being a dynamically old comet as defined by \citet{2017MNRAS.472.4634K} with a mature nucleus, depleted in volatile content, the actual context could be far more complex. There are only 28 known comets with $e > 1.0$ and $q$ inside the perihelion of Mercury, C/2021~O3 is one of them. Most of these objects were bright enough to be considered ``Great Comets" --- consider the case of C/1962~C1 (Seki-Lines), for example --- but others failed to reach such a level of brightness. A contemporary example of an object orbitally similar to C/2021~O3 that failed to achieve such status is C/2019~Y4 (ATLAS) that in April 2020 experienced a chain of breakup events that led to the formation of dozens of fragments \citep{2020AJ....160...91H,2021AJ....162...70Y}. \citet{2020AJ....160...91H} have argued that this comet could be the byproduct of the splitting of a larger cometary progenitor during its previous perihelion. Given the caveats of the interpretation of the results of numerical simulations based on contemporary data discussed at the end of Sect.~\ref{evolution}, C/2021~O3 could be a somewhat dynamically new comet masquerading as a dynamically old one if it is a piece of a dynamically new comet that fragmented during its first perihelion, perhaps tens of thousands of years ago.
    
    \section*{Acknowledgements}

    We thank the reviewer, Julio A. Fernandez, for a timely and thought-provoking report that contributed to considering alternative explanations for our results. MES acknowledges funding through a CAPES PhD fellowship, and a fellowship of the PCI Program of the Ministry of Science, Technology and Innovation, financed by the Brazilian National Council of Research - CNPq. JMC acknowledges funding through a CNPq fellowship. Support by CNPq (310964/2020-2) and FAPERJ (E-26/201.001/2021) is acknowledged by DL. JM and WP acknowledge CNPq and CAPES, respectively, for their PhD fellowship. ER and FM acknowledge FAPERJ fellowships, processes E-26/204.602/2021 and E-26/201.877/2020. PA acknowledges PCI/CNPq for  support through a fellowship.
    The authors are grateful to the IMPACTON team, in special to R. Souza, A. Santiago and J. Silva for the technical support. Observations were obtained at the Observatório Astronômico do Sertão de Itaparica (OASI, Itacuruba) of the Observatório Nacional (ON-Brazil) and observations obtained at the Southern Astrophysical Research (SOAR) telescope, which is a joint project of the Minist\'{e}rio da Ci\^{e}ncia, Tecnologia e Inova\c{c}\~{o}es (MCTI/LNA) do Brasil, the US National Science Foundation’s NOIRLab, the University of North Carolina at Chapel Hill (UNC), and Michigan State University (MSU). RdlFM and CdlFM thank S.~J. Aarseth for providing one of the codes used in this research and A.~I. G\'omez de Castro for providing access to computing facilities. Part of the calculations and the data analysis were completed on the Brigit HPC server of the `Universidad Complutense de Madrid' (UCM), and we thank S. Cano Als\'ua for his help during this stage. This work was partially supported by the Spanish `Agencia Estatal de Investigaci\'on (Ministerio de Ciencia e Innovaci\'on)' under grant PID2020-116726RB-I00 /AEI/10.13039/501100011033.
    
\section*{Data Availability}
The data underlying this article will be shared on reasonable request to the corresponding author.



\bibliographystyle{mnras}
\bibliography{C2021O3} 








\bsp	
\label{lastpage}
\end{document}